# Quantum Bit Regeneration

Isaac L. Chuang *and* Yoshihisa Yamamoto

*ERATO Quantum Fluctuation Project*
*Edward L. Ginzton Laboratory, Stanford University, Stanford, CA 94305*

(January 4, 1996)

Decoherence and loss will limit the practicality of quantum cryptography and computing unless successful error correction techniques are developed. To this end, we have discovered a new scheme for perfectly detecting and rejecting the error caused by loss (amplitude damping to a reservoir at $T=0$), based on using a dual-rail representation of a quantum bit. This is possible because (1) balanced loss does not perform a "which-path" measurement in an interferometer, and (2) balanced quantum nondemolition measurement of the "total" photon number can be used to detect loss-induced quantum jumps without disturbing the quantum coherence essential to the quantum bit. Our results are immediately applicable to optical quantum computers using single photonics devices.

42.50.Ar,89.80.th,42.79.Ta,03.65.Bz

Essential to the success of quantum cryptography and computing is the ability to create a quantum bit (qubit) and to maintain its fragile superposition state for long periods of time. A key ingredient will be the development of simple and effective quantum error correction schemes. One particularly important classical technique is regeneration, in which periodic measurement and reconstruction is used to prevent multiplicative (exponential) growth of errors and thus preserve signal integrity. However, application of the analogous procedure to a qubit is not straightforward, because no more than one bit of information can be extracted from a two-state system; simple measurement collapses the wavefunction, causing loss of information about the qubit's superposition state. Classical and quantum regeneration are similar in that redundancy must be introduced in order to allow for error correction, but different in that quantum regeneration must be performed without actually measuring the qubit being transmitted.

We have discovered a very simple scheme for quantum regeneration, under certain circumstances, which is made possible by two key insights: (1) balanced loss in the two arms does not perform a "which-path" measurement in an interferometer, and (2) balanced quantum nondemolition (QND) measurement of the "total" photon number can be used to determine whether quantum jumps due to loss have occurred, while preserving the essential linear superposition state. More specifically, by using a "dual-rail" encoding [1] of the logical zero and one qubit states as $|01\rangle$ and $|10\rangle$, we can take advantage of the fact that equal loss will always either leave the state intact or cause a jump to the $|00\rangle$ state. Such jumps can be detected using a balanced QND measurement of the total photon number (not an ordinary QND measurement of the photon number in a single mode). We explain this in detail below.

Consider the classical interferometer shown in Figure 1. It is well known that to achieve maximum fringe visibility at the output, it is necessary for the loss in both arms to be equal; furthermore, despite the loss, unit visibility can be achieved. This can be easily seen as follows: let the interferometer inputs be $E_0^a = A\cos\omega t$ and $E_0^{\bar{a}} = 0$. For 50/50 beamsplitters, the field in the arms will then be $E_1^a = E_1^{\bar{a}} = A\cos\omega t/\sqrt{2}$. Equal loss $\gamma$ in both arms causes the state immediately before the final beamsplitter to be $E_2^a = E_2^{\bar{a}} = Ae^{-\gamma}\cos\omega t/\sqrt{2}$, but despite this, the final output state is $E_3^a = Ae^{-\gamma}\cos\omega t$ and $E_3^{\bar{a}} = 0$. Moreover, the visibility is given by extremizing the output intensities over a variable phase delay inserted in one arm, and in this case, we find that $V = \left[|E_3^a|^2 - |E_3^{\bar{a}}|^2\right] / \left[|E_3^a|^2 + |E_3^{\bar{a}}|^2\right] = 1$, which is ideal. Equal loss in both arms leads to no decrease in visibility.

The same applies in the quantum interferometer when we use a single photon. Let the input be the one photon state $|\psi_0\rangle = |10\rangle$, where the two labels give the state of modes $a$ and $\bar{a}$. For beamsplitters of angle $\theta = \tan^{-1}(c_1/c_0)$, the state in the arms is $|\psi_1\rangle = c_0|01\rangle + c_1|10\rangle$. Equal loss in both arms causes the state immediately before the final beamsplitter to be

$$|\psi_2\rangle = \begin{cases} c_0|01\rangle + c_1|10\rangle & \text{with probability } e^{-\gamma} \\ |00\rangle & \text{with probability } (1-e^{-\gamma}) \end{cases}. \quad (1)$$

Letting $\alpha = |c_0|^2$ and $\beta = c_0 c_1^*$, we find the corresponding density matrix to be

$$\rho_2 = \$_\Gamma |\psi_1\rangle\langle\psi_1| = \begin{bmatrix} 1-e^{-\gamma} & 0 & 0 & 0 \\ 0 & \alpha e^{-\gamma} & \beta e^{-\gamma} & 0 \\ 0 & \beta^* e^{-\gamma} & (1-\alpha)e^{-\gamma} & 0 \\ 0 & 0 & 0 & 0 \end{bmatrix} \begin{matrix} |00\rangle \\ |01\rangle \\ |10\rangle \\ |11\rangle \end{matrix}, \quad (2)$$

where the basis states are given on the right. $\$_\Gamma = \$_\Gamma^a \$_\Gamma^{\bar{a}}$, where $\$_\Gamma^a$ is a superscattering operator acting on mode $a$ defined by

$$\$_\Gamma^a |0\rangle\langle 0| = |0\rangle\langle 0| \quad (3)$$
$$\$_\Gamma^a |0\rangle\langle 1| = e^{-\gamma/2}|0\rangle\langle 1| \quad (4)$$

$$\$_\Gamma^a |1\rangle\langle 0| = e^{-\gamma/2} |1\rangle\langle 0| \qquad (5)$$
$$\$_\Gamma^a |1\rangle\langle 1| = e^{-\gamma} |1\rangle\langle 1| + (1 - e^{-\gamma}) |0\rangle\langle 0|, \qquad (6)$$

and $\$_\Gamma^{\overline{a}}$ follows similarly. These are obtained from the usual density matrix approach [2] for amplitude damping to a reservoir at absolute zero in the Born-Markov approximation, with the interaction Hamiltonian

$$H = \gamma' \sum_k \left( a^\dagger c_k + c_k^\dagger a \right) + \gamma' \sum_k \left( \overline{a}^\dagger d_k + d_k^\dagger \overline{a} \right), \qquad (7)$$

where $c$ and $d$ are reservoir operators. Alternatively, the quantum Monte-Carlo wavefunction technique [3–5] provides a picture of the evolution of a single wavefunction. The result well describes the physical situation experienced by single photons traversing an optical fiber, where scattering is the main cause of errors and phase decoherence is negligible. The decay of the diagonal terms corresponds directly to loss of probability amplitude for finding a photon in one of the arms, while the decay of the off-diagonals is usually associated with "decoherence." Although the latter is true for damping of the usual $|0\rangle$ and $|1\rangle$ representation of a qubit, it is not valid in our dual-rail qubit case. Here, coherence between the $|01\rangle$ and $|10\rangle$ states is actually *preserved* when no quantum jump occurs because of the symmetry of the damping; $|01\rangle$ and $|10\rangle$ suffer identically under $\$_\Gamma$.

The final state is given by taking the inverse beamsplitter transform of the above, which gives

$$|\psi_3\rangle = \begin{cases} |10\rangle & \text{with probability } e^{-\gamma} \\ |00\rangle & \text{with probability } (1 - e^{-\gamma}) \end{cases}, \qquad (8)$$

since in the ideal case the second 50/50 beamsplitter simply undoes the action of the first, and otherwise it does nothing to the vacuum state $|00\rangle$. If we throw out those cases in which no photon is registered by either of the two output counters, then we find that the visibility is ideal, just as in the classical interferometer with balanced loss.

Suppose now that we stretch out the interferometer such that the middle section extends for many kilometers. Along this transmission link, loss causes quantum jumps which result in $|00\rangle$ states. How may we discriminate this state from $c_0 |01\rangle + c_1 |10\rangle$ for arbitrary $c_0$ and $c_1$? The solution is a "balanced" QND measurement of the *total photon number*. For example, we may envision the quantum circuit shown in Figure 2, where two Kerr media are used to cross-phase modulate a probe signal. When either arm contains a photon, the probe receives a $\pi$ phase shift; if neither or both arms contain a photon, the probe receives no phase shift. From Eq.(1), we have that

$$|\psi_0\rangle = \begin{cases} c_0 |0110\rangle + c_1 |1010\rangle \\ |0010\rangle \end{cases}, \qquad (9)$$

using the labeling $|a\overline{a}b\overline{b}\rangle$. This is a mixed state, with the probability of the upper and lower states being $e^{-\gamma}$ and $1 - e^{-\gamma}$, respectively. The first 50/50 beamsplitter gives us

$$|\psi_1\rangle = \begin{cases} \frac{c_0}{\sqrt{2}} \left[ |0110\rangle + |0101\rangle \right] + \frac{c_1}{\sqrt{2}} \left[ |1010\rangle + |1001\rangle \right] \\ \frac{1}{\sqrt{2}} \left[ |0010\rangle + |0001\rangle \right] \end{cases}, \qquad (10)$$

which is followed by the two Kerr media,

$$|\psi_2\rangle = \begin{cases} \frac{c_0}{\sqrt{2}} \left[ -|0110\rangle + |0101\rangle \right] + \frac{c_1}{\sqrt{2}} \left[ -|1010\rangle + |1001\rangle \right] \\ \frac{1}{\sqrt{2}} \left[ |0010\rangle + |0001\rangle \right] \end{cases}, \qquad (11)$$

and then the second beamsplitter, to give the output

$$|\psi_3\rangle = \begin{cases} c_0 |0101\rangle + c_1 |1001\rangle \\ |0010\rangle \end{cases}. \qquad (12)$$

The final measurement allows us to select the $b\overline{b} = 01$ probe state, such that the transmitted qubit is guaranteed to be

$$|\psi_{\text{out}}\rangle = c_0 |01\rangle + c_1 |10\rangle \qquad (13)$$

with probability $e^{-\gamma}$. In analogy to the quantum-optical Fredkin gate [6,7], a $\pi$ phase shift unbalances the probe interferometer, switching the output and thus discriminating the $\{|01\rangle, |10\rangle\}$ manifold perfectly from the $|00\rangle$ state. Since only total photon number information is obtained, this is a QND measurement, and the back-action is a randomization of the phase between the $\{|01\rangle, |10\rangle\}$ manifold and the $|00\rangle$ state; however, the phase coherence between the $|01\rangle$ and $|10\rangle$ states is left intact because the measurement does not discriminate between them. Specifically, the QND observable [8] is $Q = a^\dagger a + \overline{a}^\dagger \overline{a}$, and the state $|\phi\rangle = c_0 |01\rangle + c_1 |10\rangle$ is an eigenstate of the QND observable, i.e., $Q|\phi\rangle = |\phi\rangle$; thus, the linear superposition state $|\phi\rangle$ is projected out by this QND measurement.

This device is an ideal regenerator in the following sense: (1) it detects perfectly when an error occurs, by discriminating illegal states without destroying a legal wavefunction, and (2) it can prevent multiplicative growth of error. Although the latter is not true when loss is exponential (as for linear loss in fibers), error growth *can* be prevented when loss occurs at a sub-exponential rate. Suppose that instead of $e^{-\gamma}$ we have the loss $1 - \epsilon t^2$ after time $t$, for small $\epsilon$; this may be the case, for example, for spontaneous emission by cavity confined atoms. Without regeneration, the final output is correct with probability $1 - \epsilon n^2$ after $n$ steps; however, when regeneration is performed after each step, the probability of a correct result is $(1 - \epsilon)^n \approx 1 - n\epsilon$, which is much better. This result is known as the watchdog effect [9], and is purely a quantum-mechanical effect; in fact, by regenerating infinitely often, evolution is suspended entirely by virtue of the quantum zeno effect, and amplitude damping is prohibited from happening.

Our results suggest the following scheme for transmission of a quantum bit: the two states $|01\rangle$ and $|10\rangle$ are used as basis states to form the arbitrary qubit $c_0|01\rangle + c_1|10\rangle$. Physically, this may be generated using a single photon incident on a beamsplitter and a phase shifter. Under normal operation, the state satisfies the *representation invariant* condition $a^\dagger a + \overline{a}^\dagger \overline{a} = 1$, but when quantum jumps due to loss occurs, the illegal state $|00\rangle$ results. This is true only when both modes suffer equal loss, but that may be guaranteed experimentally by using time-multiplexing to send both modes down the same optical fiber. To regenerate, we discriminate $|00\rangle$ from the representation manifold spanned by $|01\rangle$ and $|10\rangle$ by using a balanced QND measurement of the total photon number, which indicates if an error has occurred or not without introducing back-action noise into the representation manifold. If an error occurs, we abort the transmission and request the sender to try again. For exponential loss, $e^\gamma \sim 1 + n^2\epsilon$ trials are required to transmit a perfect qubit, but for sub-exponential error probability $1 - \epsilon$ per step, only approximately $1 + n\epsilon$ trials are required with periodic regeneration. Perhaps the most interesting point is that this scheme provides *error-free* transmission, in contrast to classical regeneration, which requires acceptance of a finite error probability.

Classical information theory describes a close analogy to our system: the binary erasure channel [10], in which 1 and 0 are transmitted perfectly with probability $e^{-\gamma} = \alpha$, and otherwise an error symbol $e$ is received. This is an elementary model which describes the effect of classical noise due to loss, similar to the noise due to the super-scattering operator $\$_\Gamma$. However, there is a subtle and important distinction that must be made: the capacity of the classical channel is $\alpha$ bits. In contrast, according to our result, the capacity of the quantum channel is at least $\alpha/2$ *qubits* (the factor of two comes from our use of two qubits to code each dual-rail qubit). A quantum bit is different from a classical one; the receiver obtains not only the diagonal elements $|c_0|^2$ and $|c_1|^2$, but also the off-diagonals $c_0 c_1^*$ and $c_0^* c_1$, which may communicate information about entanglement with other states. Shannon's noisy coding theorem defines the capacity of a noisy classical channel; the equivalent for quantum channels is presently unknown [11].

Practically speaking, we anticipate that our scheme may be useful to quantum cryptography, where it is necessary to guarantee the integrity of transmitted qubits, but repeated transmission is allowed since it is permissible to change the qubit sent each time re-transmission is required. Furthermore, it is simple to show that cross-phase modulators, beamsplitters, and phase shifters form a complete set of operations necessary to perform logic with dual-rail qubits, and thus our scheme is directly applicable to quantum computation. For example, it may be applied to correct loss induced errors in the single-photonics quantum computation proposal of [1]. The crucial impediment is the realization of a Kerr medium with sufficiently strong nonlinearity to obtain $\pi$ cross-phase modulation between single photons; the good news is that recent experimental results indicate that resonant effects in atomic [12] and excitonic [13] cavity QED may provide the key. Our scheme may also be used to correct errors due to spontaneous emission in ion trap quantum computers [14]; pairs of ions or states within ions can be used as dual-rail qubits, with regeneration being performed using cross-phase modulation with a probe quantum bit via the center-of-mass phonon mode "bus" qubit.

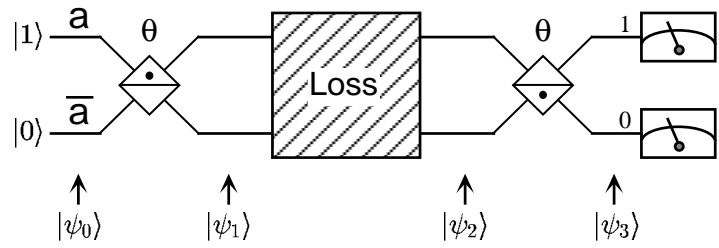

FIG. 1. Classical interferometer with equal loss in both arms (modes $a$ and $\bar{a}$). The two beamsplitters are inverses of each other. The expected results are shown to the left of the meters at the outputs.

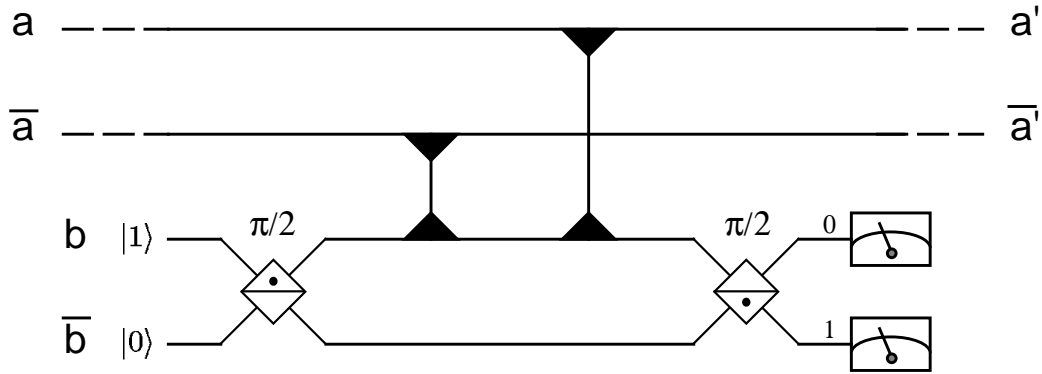

FIG. 2. Quantum optical regenerator for dual-rail qubits using a balanced QND measurement of the total photon number. The top two wires carry the transmitted qubit, and the bottom two the probe. Triangles connected by vertical lines represent $\pi$ cross-phase shift Kerr media. The input is on the left and the output to the right.